\documentclass[conference]{IEEEtran}

\usepackage[noadjust]{cite}
\usepackage{amsmath,amssymb,amsfonts}
\usepackage{algorithmic}
\usepackage{graphicx}
\usepackage{textcomp}
\usepackage[dvipsnames]{xcolor}

\usepackage{url}

\usepackage{amsthm}

\usepackage{xspace}

\usepackage{relsize}

\theoremstyle{definition}

\newcommand{\EE}{\mathbb{E}}

\newcommand{\RR}{\mathbb{R}}

\newcommand{\mcL}{\mathcal{L}}

\usepackage{cleveref}

\title{Error Analysis of Option Pricing via Deep PDE Solvers: Empirical Study}

\author{\IEEEauthorblockN{Rawin Assabumrungrat}
\IEEEauthorblockA{\textit{School of Engineering} \\
\textit{Tohoku University}\\
Sendai, Japan \\
rawinassa@gmail.com
}
\and
\IEEEauthorblockN{Kentaro Minami}
\IEEEauthorblockA{\textit{Preferred Networks, Inc.}\\
Tokyo, Japan \\
}
\and
\IEEEauthorblockN{Masanori Hirano}
\IEEEauthorblockA{\textit{Preferred Networks, Inc.}\\
Tokyo, Japan \\
research@mhirano.jp
}
}

\begin{document}

\maketitle

\begin{abstract}
Option pricing, a fundamental problem in finance, often requires solving non-linear partial differential equations (PDEs). When dealing with multi-asset options, such as rainbow options, these PDEs become high-dimensional, leading to challenges posed by the curse of dimensionality. While deep learning-based PDE solvers have recently emerged as scalable solutions to this high-dimensional problem, their empirical and quantitative accuracy remains not well-understood, hindering their real-world applicability. In this study, we aimed to offer actionable insights into the utility of Deep PDE solvers for practical option pricing implementation. Through comparative experiments, we assessed the empirical performance of these solvers in high-dimensional contexts. Our investigation identified three primary sources of errors in Deep PDE solvers: (i) errors inherent in the specifications of the target option and underlying assets, (ii) errors originating from the asset model simulation methods, and (iii) errors stemming from the neural network training. Through ablation studies, we evaluated the individual impact of each error source. Our results indicate that the Deep BSDE method (DBSDE) is superior in performance and exhibits robustness against variations in option specifications. In contrast, some other methods are overly sensitive to option specifications, such as time to expiration. We also find that the performance of these methods improves inversely proportional to the square root of batch size and the number of time steps. This observation can aid in estimating computational resources for achieving desired accuracies with Deep PDE solvers. 
\end{abstract}


\section{Introduction}

Option pricing is a crucial problem in finance.
One of the fundamental tools in option pricing is the pricing framework based on risk-neutral measures \cite{shreve2004stochastic2,Bingham_Kiesel_2004}.
The calculation of risk-neutral prices has been frequently formulated as a solution to partial differential equations (PDEs) since the seminal contribution of Black and Scholes (1973) \cite{black1973pricing}.
The requirements of practical finance encompass the valuation of options on multiple underlying assets, such as basket options or credit value adjustments (CVAs), which entail the possibility of high-dimensional PDEs.
However, conventional techniques, including finite difference methods and finite element methods, fall prey to the curse of dimensionality, posing a challenge to their scalability to high-dimensional PDEs.
Therefore, attention has recently been focused on new PDE solvers that can scale to high dimensions \cite{e2021review}.

Deep BSDE solver (\cite{han2018solving,e2017deep}) is a promising high-dimensional PDE solver based on deep learning, which solves the equivalent backward stochastic differential equation (BSDE) of the PDE to be solved by optimizing the function modeled by a neural network.
See \cite{beck2020overview,germain2021neural,e2021review,chessari2023survey} for comprehensive reviews of deep PDE/BSDE solvers and their variants.

Practical option pricing demands a high level of accuracy.
However, the current challenge of deep PDE/BSDE solvers is the lack of quantitative accuracy guarantees to achieve the required precision.
One reason for this is that current neural network-based function approximation methods commonly incorporate many heuristics, and thus practitioners utilize them as black-box methods.
To date, several researchers have investigated the theoretical approximation accuracy of deep learning-based PDE solvers (\cite{grohs2020proof,hutzenthaler2020proof}).
Nevertheless, the existing theoretical analyses are mainly centered on the existence of neural networks that approximate solutions.
As such, conducting empirical research is imperative to gain a comprehensive understanding of the accuracy of deep PDE methods with practical implementations of neural network models and algorithms.

Our primary objective is to provide practical guidelines for the design of models and training algorithms in option pricing using Deep PDE/BSDE solvers.
To this end, it is essential to deeply understand the errors in these methods.
There can be multiple sources of errors, including
(i) uncertainties inherent in options and their underlying assets, such as high volatility or long maturity,
(ii) errors related to asset models, such as discretization errors in simulations, and
(iii) optimization errors related to the choice of neural network architectures and optimization algorithms.
We elaborate on these issues in \cref{sec:source}.

In this paper, we empirically investigate the performance of Deep PDE/BSDE solvers in high-dimensional option pricing.
In our experiments in \cref{sec:experiment}, we examine the performance of existing Deep PDE solvers for pricing high-dimensional rainbow options and discuss the influence of each source of error.
Notably, our results demonstrate that the Deep BSDE method \cite{e2017deep,han2018solving} consistently outperforms comparative methods, showcasing a consistent robustness to fluctuating option parameters, e.g., time to expiration.
Additionally, performance metrics for these methods appear to improve in inverse proportion to the square root of the batch size (proportion to $1/\sqrt{\text{batch size}}$) and the number of time discretization steps (proportion to $1/\sqrt{\text{time steps}}$), which might be useful for computational budgeting for achieving accuracy required for financial applications.

\section{Background}\label{sec:background}

\subsection{Option pricing via PDE and BSDE}
Here, we will review the basics of option pricing using PDEs.
We will omit a comprehensive and mathematically rigorous formulation as it is beyond the scope of this paper.
See e.g.~\cite{shreve2004stochastic2} for an introduction to this topic.

An option price is often formulated as a conditional expectation with respect to a stochastic process.
Let $X_t$ ($t \geq 0$) be a $d$-dimensional process of the underlying asset prices.
The asset price dynamics is typically modeled by a stochastic differential equation (SDE)
\begin{equation}
    d X_t = \mu(t, X_t) dt + \sigma(t, X_t) dW_t,
    \label{eq:forward-sde}
\end{equation}
where $W_t$ is a $d$-dimensional Wiener process, $\mu$ is an $\RR^d$-valued function, and $\sigma$ is a $d \times d$ matrix-valued function.
More precisely, we consider the dynamics of $X_t$ under the risk-neutral measure,
where the expected returns of the risky assets are adjusted to match the risk-free rate $r$.
We are interested in pricing an option with maturity $T > 0$.
The payoff of the option depends on the price process $X_t$.
For example, a European-type option has a payoff $\phi(X_T)$ that depends on the terminal value of the underlying assets.
Given an initial condition $X_t = x$, the price of the option at time $t$ is given as a conditional expectation
$
    u(t, x) = \EE[ \exp(-r (T - t)) \phi(X_T) \mid X_t = x].
$

The option price $u(t, x)$ can be linked to a PDE as follows.
Let $\mcL$ be a second-order differential operator defined as
\[
    \mcL = \sum_{i} \mu_i(t, x) \frac{\partial}{\partial x_i}
    + \frac{1}{2} \sum_{i, j} [\sigma \sigma^T]_{i, j}(t, x) 
    \frac{\partial^2}{\partial x_i \partial x_j}.
\]
Then, by the Feynman--Kac formula, $u(t, x)$ solves the Black--Scholes PDE \cite{black1973pricing}
\begin{equation*}
    \frac{\partial}{\partial t} u(t, x) + \mcL u(t, x) = r u(t, x),
    \quad t \in [0, T], x \in \RR^d
\end{equation*}
with a terminal condition $u(T, x) = \phi(x)$, $x \in \RR^d$.
For more intricate options (e.g., exotics), the payoff can depend on the entire trajectory of $X_t$, but their prices can still be formulated as solutions of PDEs (see, e.g., \cite[Chap.~7]{shreve2004stochastic2}).
More generally, this formulation can be extended to a parabolic PDE
\begin{equation}
    \frac{\partial}{\partial t} u(t, x) + \mcL u(t, x)
    = f(t, x, u(t, x), \sigma^\top(t, x) \nabla_x u(t, x)).
    \label{eq:semilinear-pde}
\end{equation}
with a non-linear contribution term $f$,
which covers a wider range of financial applications, such as xVA computation (\cite{burgard2011counterparty, burgard2013funding}) and stochastic control problems.

Furthermore, the solution $u(t, x)$ of \eqref{eq:semilinear-pde} can be characterized as a solution of a BSDE \cite{pardoux1992backward}.
To this end, we define new processes $Y_t = u(t, X_t)$ and $Z_t = \nabla_x u(t, X_t)$,
where $X_t$ follows the (forward) SDE \eqref{eq:forward-sde} with the initial condition $X_{t_0} = x$.
Then, $(Y_t, Z_t)$ satisfies the following BSDE
\begin{equation}
    Y_t = \phi(X_T)
    - \int_t^T f(s, X_s, Y_s, Z_s) ds
    - \int_t^T Z_s \cdot d W_s
    \label{eq:backward-sde}
\end{equation}
for $t_0 \leq t \leq T$.
Conversely, if $(Y_t, Z_t)$ solves \eqref{eq:backward-sde}, then $Y_{t_0} = u(t_0, x)$ becomes a (viscosity) solution of the PDE \eqref{eq:semilinear-pde}.
This observation forms the crux of Deep BSDE methods discussed in the following subsection.

\subsection{Deep BSDE methods}
\label{DeepBSDEmethods}
In recent years, there has been a surge of interest in algorithms that utilize deep learning techniques for approximating solutions of PDEs (see e.g.~\cite{beck2020overview,blechschmidt2021threeways}).
The Deep BSDE method, first introduced by \cite{e2017deep,han2018solving}, leverages the aforementioned correspondence between PDEs and BSDEs to achieve scalability for high-dimensional PDEs.
Here, we provide an overview of several variants of the Deep BSDE method. Comprehensive reviews can be found in \cite{beck2020overview,germain2021neural,e2021review,chessari2023survey}.

In what follows, we use the following notations.
Let $t_0 = 0 < t_1 < \cdots < t_N = T$ be a subdivision of the time interval $[0, T]$ and $\Delta t_i := t_{i+1} - t_i$ for $i = 0, 1, \ldots, N-1$.
We consider the Euler discretization of \eqref{eq:forward-sde} defined by
\begin{equation}
    \widehat{X}_{i+1}
    = \widehat{X}_i + \mu(t_i, \widehat{X}_i) \Delta t_i
    + \sigma(t_i, \widehat{X}_i) \Delta W_i
    \label{eq:forward-euler}
\end{equation}
for $i = 0, 1, \ldots, N - 1$,
where $\hat{X}_0 = X_0$ is the initial value of the original SDE and $\Delta W_i := W_{t_{i+1}} - W_{t_i}$.
Note that $\Delta W_i$ are independent normal random variables with mean $0$ and variance $\Delta t_i$,
so we can numerically simulate \eqref{eq:forward-euler}.
Below, the expectation symbol $\EE$ are to be understood as empirical expectations based on simulations of the Euler method.

\subsubsection{Deep BSDE method (DBSDE)}

The \textit{Deep BSDE method (DBSDE)} \cite{han2018solving,e2017deep}, also known as the ``forward scheme'', leverages the forward induction representation of the BSDE \eqref{eq:backward-sde}:
\begin{align*}
    \widehat{Y}_{i+1} & = \widehat{Y}_{i+1}^{u_0, \widehat{Z}} \\
    & := \widehat{Y}_i + f(t_i, \widehat{X}_i, \widehat{Y}_i, \widehat{Z}_i(\widehat{X}_i)) \Delta t_i
    +  \widehat{Z}_i(\widehat{X}_i) \cdot \Delta W_i.
\end{align*}
To ensure that $\widehat{Y}_N$ satisfies the terminal condition, we minimize the following objective 
\[
    J(u_0, \{ \widehat{Z}_i \}_{i=0}^{N-1}) := \EE \left | \widehat{Y}_N^{u_0, \widehat{Z}} - g(\hat{X}_N) \right|^2
\]
with respect to the initial value $\hat{Y}_0 = u_0$ and functions $\widehat{Z}_i: \RR^d \to \RR^d$ ($i = 0, \ldots, N-1$), where each $\widehat{Z}_i$ is modeled by a neural network.
The output of the method is the approximate solution $u_0 = u(0, X_0)$ at time $t = 0$.

\subsubsection{Deep Backward Dynamic Programming (DBDP)}

The \textit{Deep Backward Dynamic Programming} \cite{hure2020backward} is another approach based on the following backward induction scheme
\[
    \widehat{Y}_{i}
    = \widehat{Y}_{i+1}
    - f(t_i, \widehat{X}_i, \widehat{Y}_i, \widehat{Z}_i) \Delta t_i
    - \widehat{Z}_i \cdot \Delta W_i.
\]
Starting from $U_N = g$, this method trains neural networks $\{ U_i \}_{i=0}^{N-1}$ and $\{ V_i \}_{i=0}^{N-1}$ to minimize the following objective function
\begin{align*}
    J_i(U_i, V_i)
    & := \EE \Big |
        U_{i+1}(\widehat{X}_{i+1})
        - U_i(\widehat{X}_i) \\
    & - f(t_i, \widehat{X}_i, U_i(\widehat{X}_i), V_i(\widehat{X}_i)) \Delta t_i
    - V_i(\widehat{X}_i) \cdot \Delta W_i
    \Big |^2
\end{align*}
sequentially for $i = N-1, \ldots, 0$.
Here, $U_i: \RR^d \to \RR$ and $V_i: \RR^d \to \RR^d$ are unknown functions modeled by neural networks.
It should be noted that $U_i$ corresponds to the solution of the original PDE $u(t_i, \cdot)$, while $V_i$ corresponds to its gradient $\sigma^\top(t_i, \cdot) \nabla_x u(t_i, \cdot)$.
\cite{hure2020backward} proposed two variants of the method;
\textit{DBDP1} models $U_i$s and $V_i$s with independent neural networks.
\textit{DBDP2} models $U_i$s by scaler-valued neural networks, and compute $V_i(\cdot) = \sigma(t_i, \cdot)^\top \nabla_x U_i(\cdot)$ via automatic differentiation.

\subsubsection{Deep Splitting method (DS)}

The \textit{Deep Splitting (DS)} method \cite{beck2021splitting} is similar to DBDP2 in that it is based on the backward induction and it approximates $Z_t$ using automatic differentiation.
A key difference is that the DS employs network values $U_{i+1}(\widehat{X}_{i+1})$ of previous time steps in the nonlinearity $f$:
\begin{align*}
    J_i(U_i)
    & := \EE \Big |
        U_{i+1}(\widehat{X}_{i+1})
        - U_i(\widehat{X}_i) \\
    & - f(
        t_{i},
        \widehat{X}_{i+1},
        U_{i+1}(\widehat{X}_{i+1}),
        V_{i+1}
    ) \Delta t_i
    \Big |^2
\end{align*}
where $V_{i+1} := \sigma^\top(t_{i}, \widehat{X}_{i}) \nabla_x U_{i+1}(\widehat{X}_{i+1})$
\footnote{
    Here, we introduce a slightly modified version of the method introduced by \cite{germain2022approximation}, rather than the original definition \cite{beck2021splitting}.
}.


\subsubsection{Deep Backward Multistep method (MDBDP)}

The \textit{Deep Backward Multistep Method (MDBDP)} \cite{germain2022approximation} relies on the following iterated representation of the backward induction
\begin{equation}
    \widehat{Y}_i
    = g(\widehat{X}_N)
    - \sum_{j=i}^{N-1} \left[
        f(t_j, \widehat{X}_j, \widehat{Y}_j, \widehat{Z}_j) \Delta t_j
        + \widehat{Z}_j \cdot \Delta W_j
    \right].
    \label{eq:iterated-backward}
\end{equation}
For $i = N-1, \ldots, 0$, this method minimizes the objective function
\begin{align*}
    & J_j(U_i, V_i)
    := \EE \Big | \; g(\widehat{X}_N) \\
    & - \sum_{j = i+1}^{N-1}
        \left[
            f(t_j, \widehat{X}_j, U_j(\widehat{X}_j), V_j(\widehat{X}_j)) \Delta t_j
            + V_j(\widehat{X}_j) \cdot \Delta W_j
        \right] \\
    & 
    - f(t_i, \widehat{X}_i, U_i(\widehat{X}_i), V_i(\widehat{X}_i)) \Delta t_i
    - V_j(\widehat{X}_i) \cdot \Delta W_i \\
    & - U_i(\widehat{X}_i) \; \Big |^2
\end{align*}
with respect to neural networks $U_i: \RR^d \to \RR$ and $V_i: \RR^d \to \RR^d$.
Thus, the neural network architecture trained in MDBDP is similar to that of DBDP1, but it is expected that the use of the iterated representation \eqref{eq:iterated-backward} will reduce error propagation.

\section{Source of errors in deep option pricing methods}\label{sec:source}
This section provides details of the three sources of errors to which Deep PDE/BSDE solvers are subject. Additionally, the behavior of error arising from each source, as well as how it affects the solvers' performance, is explained in order to make clear the limitations of Deep PDE/BSDE solvers. These issues are then empirically illustrated through the experiments detailed in \cref{sec:experiment}. 


\subsection{Errors inherent in options' and underlying assets' uncertainties}
Option prices represent the expected present values of their total future cash flows. In general, they are determined by the price of the underlying asset (spot price), the option's exercise price (strike price), time to expiration, risk-free rate of interest, and volatility of the underlying assets. Those factors themselves partly determine the errors caused by uncertainties \cite{wang2011implied}. This paper examined the errors involved with time to expiration, spot and strike prices, and volatility, which drive the random process $X_t$.

Time to expiration refers to the time that options come due or expire. The accuracy of the numerical approximation depends on the step sizes used in the time grid. A shorter time to expiration typically requires smaller time steps, leading to a finer grid and potentially more accurate results. In contrast, a longer time can cause a coarse grid and deteriorate accuracy \cite{pironneau2009pde}. 
Additionally, the error of approximation can increase near maturity because of numerical instability. Case in point, the approximation near maturity is subject to degrade due to a non-smooth payoff function \cite{pironneau2009pde}. 
Consequently, when the time to expiration is too small, the near-maturity instability dominates. This instability results in larger pricing errors, so smaller step sizes may also be required.

Spot and strike prices play a key role in the intrinsic value of options. Here, we refer to the ratio of spot price $S_0$ to strike price $K$ as moneyness $M$. Moneyness plays a role in the pricing error of the used model. For instance, in the Black--Scholes model, the pricing error may be greater when an option is in or out of money than it is at the money. In particular, the Black--Scholes model underprices an in-the-money option, and vice versa. The model's presumptions might not accurately reflect the dynamics of options whose current spot price is far from theirs, making them prone to uncertainty.


The greater the volatility of the underlying asset, the higher the probability that the underlying price exhibits significant fluctuations. In general, as the volatility rises, the prices of all options on that underlying asset cost higher, causing the probability of the underlying price finishing in the money is higher. On the other hand, higher volatility is also associated with higher uncertainties, which are driving causes of errors in Monte Carlo methods. According to \cite{guPricingEuropean}, using the Monte Carlo (with Jump-Diffusion) model, pricing appears to be less accurate for high-volatility stocks, particularly in long-maturity options. Since a higher volatility means that the sample paths can deviate more at each time step, the computed average might not accurately reflect the distribution of underlying price in reality, which in turn jeopardizes accuracy.

\subsection{Errors by asset models}
Pricing errors in the underlying asset largely affect determined option prices \cite{louis2021}. This subsection specifies errors caused by the nature of the underlying asset's price sampling, which is an essential step to solve a PDE/BSDE using the deep learning models investigated in this paper. 

Time discretization in asset price sampling contributes to an error. This error is principally contributed by the number of time steps as well as the choice of discretization methods, e.g., the Euler method, the first and second Milstein methods, and the extrapolation method. In this paper, we limit to investigating only equal-time discretization --- that is, to define $\Delta t = T/N$ where $T$ is time to maturity and $N$ is the number of time steps. We speculate that the pricing error caused by time discretization abides by $O(1/\sqrt{N})$ \cite{dongMonteCarlo}.

The fact that the batch size is limited underscores how the sampling may not accurately represent the true distribution of the price. The estimate's standard error is proportional to $1/\sqrt{N}$, where $N$ is the total number of trajectory measurements \cite{dongMonteCarlo}. Additionally, it has been revealed that Monte Carlo methods can misestimate option prices in comparison to analytic solutions because a simulation model may, for example, underplay the tails of the distribution of $\phi(X_T)$, which are key drivers of price \cite{pedromonte}.


\subsection{Optimization errors}
Optimization errors are intrinsic in the process of optimization itself. No optimization algorithm is guaranteed to find the true optimal solution for all problems, and that includes deep option pricing, which relies extensively on neural network optimization. Nonetheless, carefully choosing the appropriate parameters makes it possible to minimize the occurrence and impact of optimization errors. In this paper, we utilize the neural networks as proposed in \cite{han2018solving}.

Different algorithms, as reviewed in \cite{germain2022approximation}, perform the approximation of PDE solutions differently. Therefore, we can hypothesize certain algorithms' superiority under certain conditions. 


The number of iterations of all the training data is crucial to obtain an accurate solution since an insufficient number can put at risk the accuracy of Deep BSDE/PDE solvers. On the other hand, choosing an unnecessarily large number causes an excessive computational cost. Another issue is that there is no recommendation for epochs required to reach an accurate solution, which also needs to balance with the learning rate used.

\section{Numerical experiments}\label{sec:experiment}


This section details the three experiments conducted to make empirical illustrations of the three types of errors encountered by Deep PDE/BSDE solvers, as described and analyzed in \cref{sec:source}. The contents are divided into two subsections: \cref{Numexp:setting} and \cref{Numexp:result}, where each of the three experiments, corresponding to each type of error, is presented. 

\subsection{Setting}
\label{Numexp:setting}
The Heston model is adopted in the same way for all algorithms \cite{10.1093/rfs/6.2.327}. 
Here the price of the asset $S_t$, is determined by a stochastic process \cite{paulonquant}
\[
    d S_t=\mu S_t d t+\sqrt{\nu_t} S_t d W_t^S
\]
where $\nu_t$, the instantaneous variance is given by
\[
    d \nu_t=\kappa\left(\theta-\nu_t\right) d t+\xi \sqrt{\nu_t} d W_t^\nu,
\]
and $W_t^S, W_t^\nu$ are Wiener processes with correlation $\rho$.
In the \textit{standard setting} of this paper, the model parameters are assigned as follows $S_0=100$, $r=0.05$, $T=1$, $\nu_0=0.1$, $\theta=0.1$, $\rho=0$, $\kappa=2$, and $\xi=0.1$: $S_0$, the initial stock price (spot price); $r$, the risk-free interest rate; $T$, the time to the option's expiration;
$\nu_0$, the initial variance;
$\theta$, the mean rate, or long-term variance of the price, so as $t$ approaches infinity, the expected value of $v_t$ approaches $\theta$;
$\rho$, the correlation of the two Wiener processes;
$\kappa$, the rate at which $v_t$ reverts to $\theta$;
$\xi$, the volatility of volatility, or `vol of vol,' which designates the variance of $v_t$. The values of $\rho$, $\kappa$, and $\xi$ are the default parameters of options provided by \cite{10.1093/rfs/6.2.327}. It is certain that those parameters satisfy the following condition, ensuring that $\nu_t$ is strictly positive: $2 \kappa \theta > \xi^2$. 

Our standard setting uses a call, best-of option of $N=20$ underlying assets. That is, the payoff function is defined by
\begin{equation}
    \phi(X_T)=\max \left\{\max _{1 \leq i \leq N} S_i-K, 0\right\}.
    \label{eq:terminalcondition}
\end{equation}
Note that $K$ stands for the strike price, hereafter defined as $S_0 / M$, and the standard moneyness $M$ is 1.2.

In addition, regarding the standard hyperparameters applied across all tested algorithms, the neural networks have a total of 4 layers, where each hidden layer has 128 nodes. The batch size used during training is set to 64, while the size of the validation set is 2048. The learning rate is 0.01. The time duration is equally divided into 40 time steps. Additionally, in the forward scheme algorithm DBSDE, a total of 8000 iterations are applied after the acceleration scheme \cite{naito2020acceleration}. In the backward scheme, the number of iterations of the initially trained time step is 16000, followed by multiple sets of 3000 iterations for the rest of the time steps. 

\subsubsection{Errors inherent in options' and underlying assets' uncertainties}

To explore the errors of this type, one of the time to expiration $T$, moneyness $M$, and mean rate $\theta$ is varied from the standard setting. The list of used time to expiration is 3/12, 6/12, 9/12, 12/12 (standard), 15/12, 18/12, 21/12 years. The moneyness values are 0.9, 1, 1.1, 1.2 (the standard), and 1.3. Lastly, the following mean rate values are applied: 0.06, 0.08, 0.10 (standard), 0.12, and 0.14. 

\subsubsection{Errors by asset models}

In this type of error, the time step division is varied to 5, 10, 20, 40 (standard), and 80 while every interval is maintained equal to each other. Alternatively, the training batch size is adjusted from 4, 16, 64 (standard) to 256. 

\subsubsection{Optimization errors}

Lastly, we investigate the optimization errors related to the choice of neural network optimization by varying the epochs. In the forward scheme, 125, 500, 2000, 8000 (standard), and 12000 iterations are applied. On the other hand, the first iterations of other backward schemes are modified to 250, 1000, 4000, 16000 (standard), and 24000. In all of the mentioned three experiments, only one of the parameters is adjusted. In other words, all of the other parameters are set to the standard setting. 

\subsection{Results}
\label{Numexp:result}

\begin{figure}[t]
 \centering
  \includegraphics[width=8.5cm]{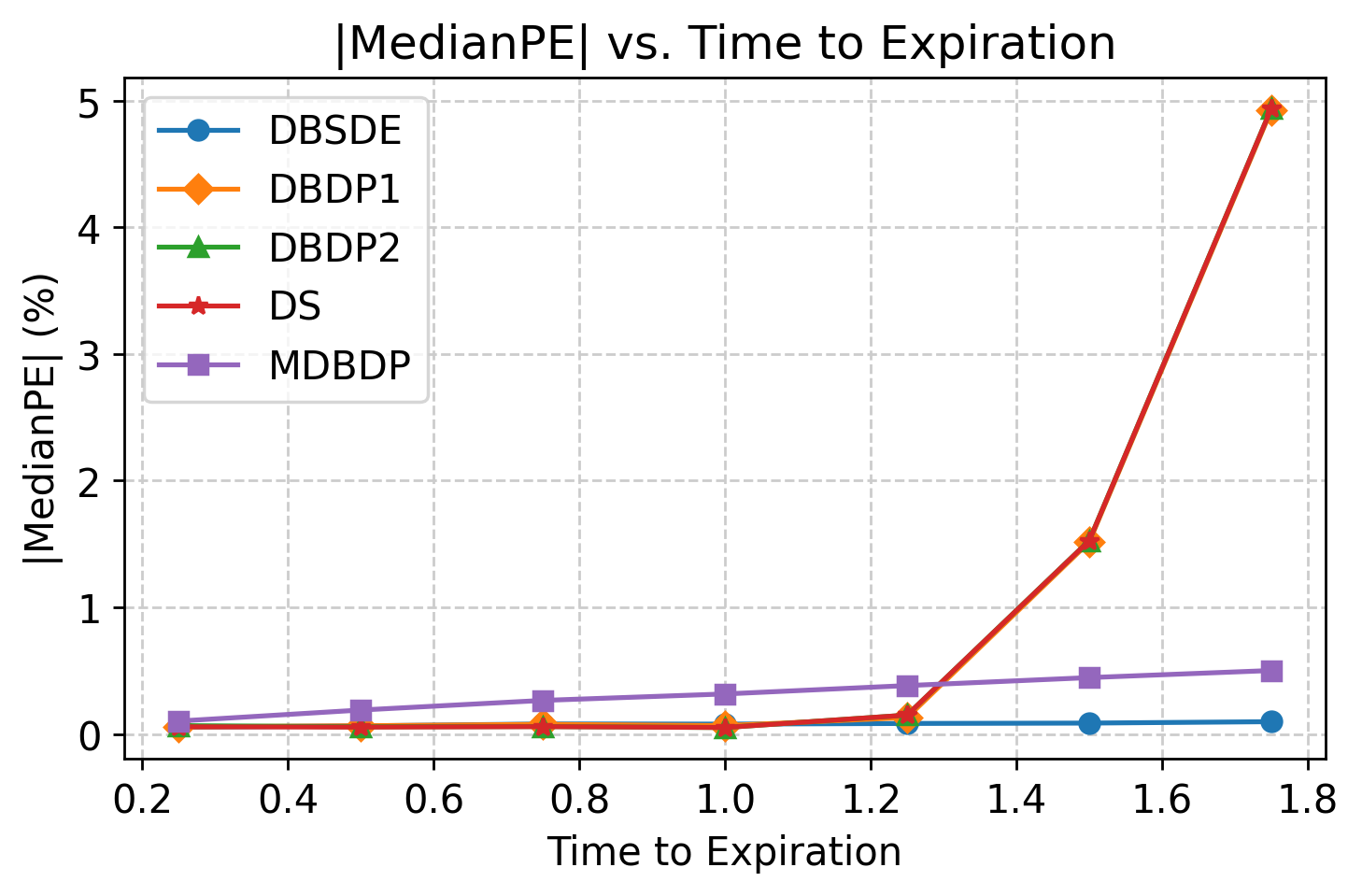}
  \vspace*{-4mm}
  \caption{Experiment 1 (Time to expiration). Both DBSDE and MDBDP exhibit a gradual increase, while DBDP1, DBDP2, and DS demonstrate a rapid ascent. Note that the trajectories of DBDP1, DBDP2, and DS almost overlap.}
  \label{fig:TvM}
\end{figure}

\begin{figure}[t]
 \centering
  \includegraphics[width=8.5cm]{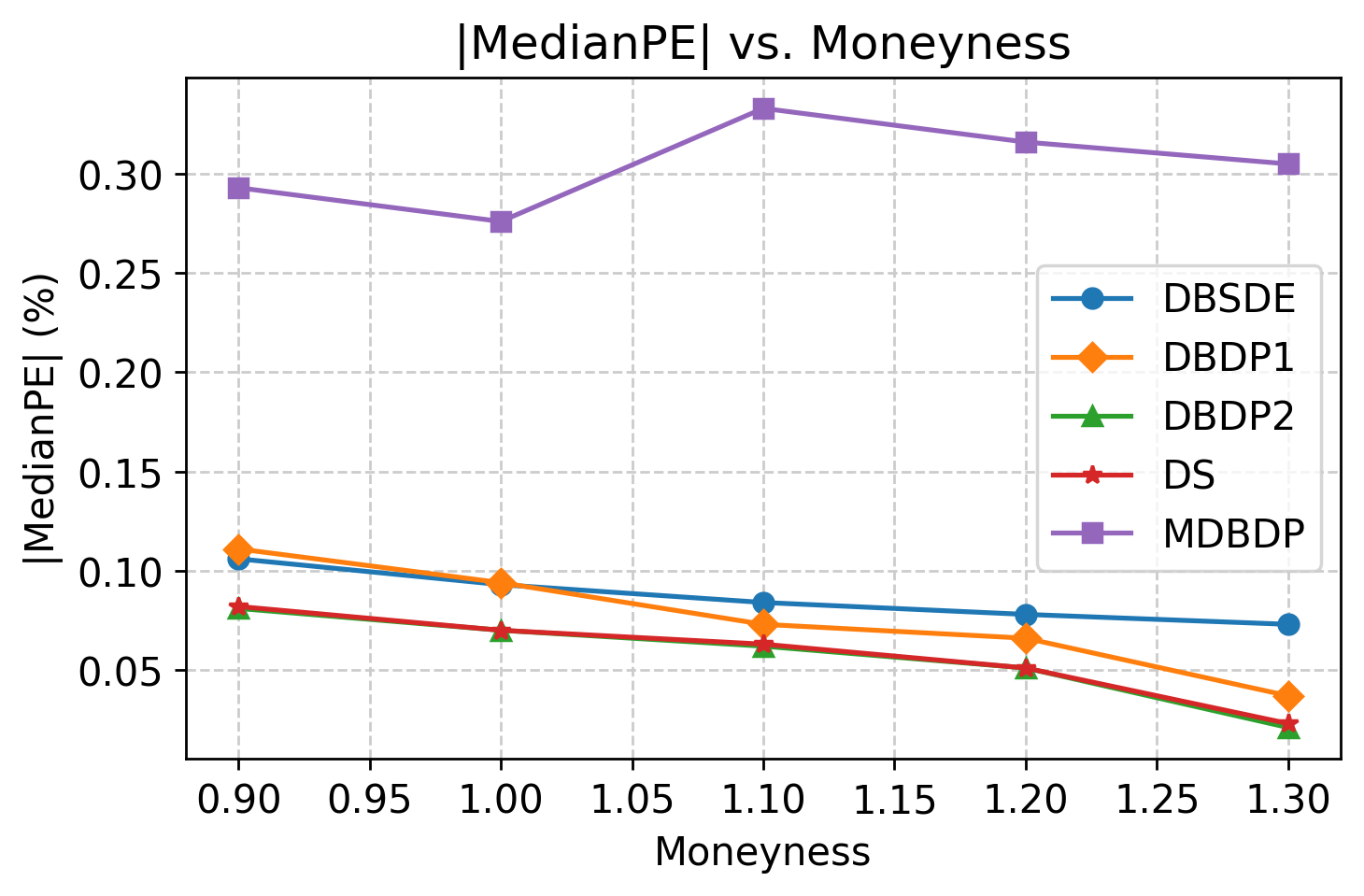}
  \vspace*{-4mm}
  \caption{Experiment 1 (Moneyness). It is obvious that MDBDP produces higher errors than the other algorithms. In this figure, DBDP2 and DS almost overlap. }
  \label{fig:MvM}
\end{figure}

\begin{figure}[t]
 \centering
  \includegraphics[width=8.5cm]{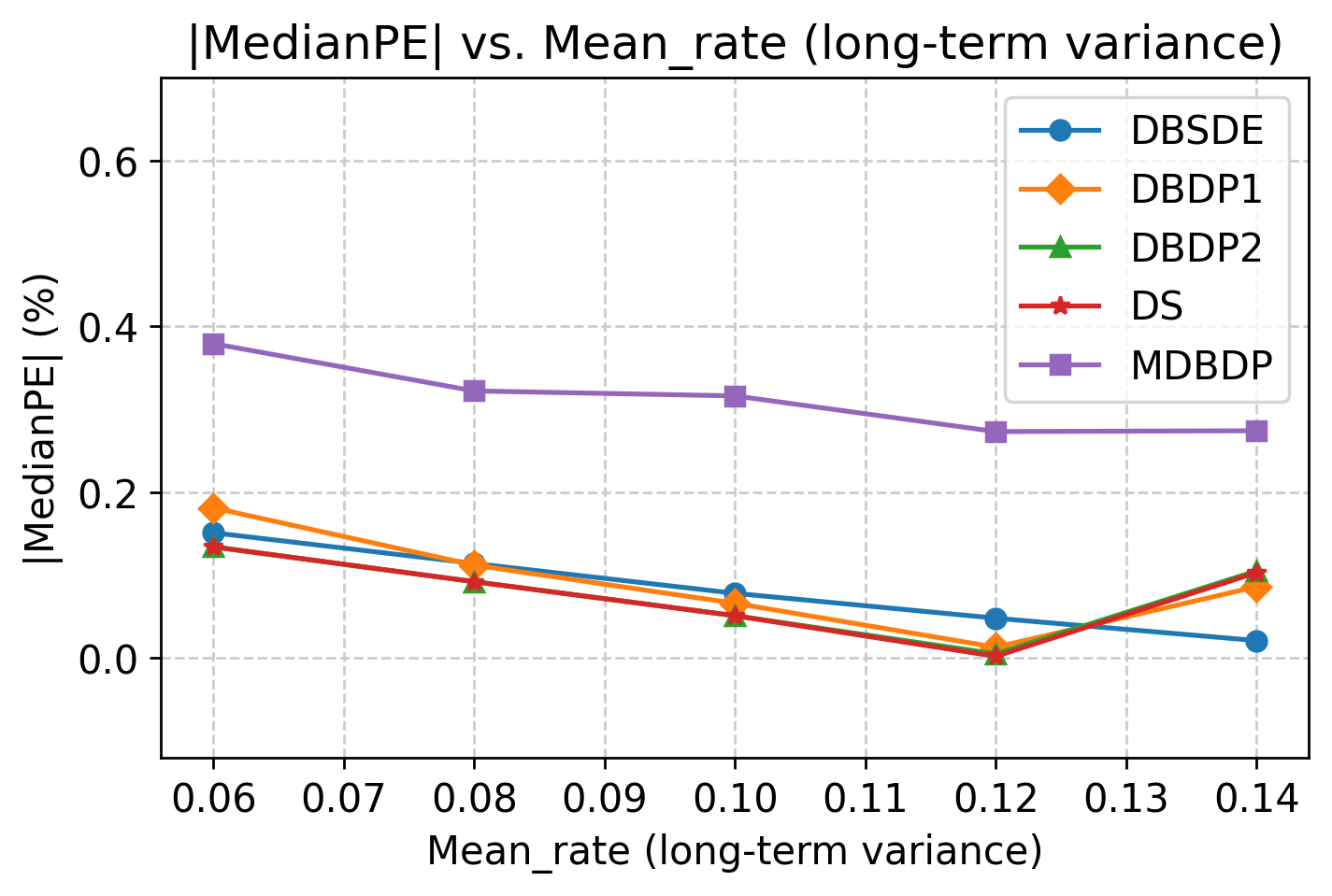}
  \vspace*{-4mm}
  \caption{Experiment 1 (Long-term variance). Similar to \cref{fig:MvM}, MDBDP produces higher errors than the other algorithms, and DBDP2 and DS almost overlap. }
  \label{fig:vvM}
\end{figure}

\begin{figure}[t]
 \centering
  \includegraphics[width=8.5cm]{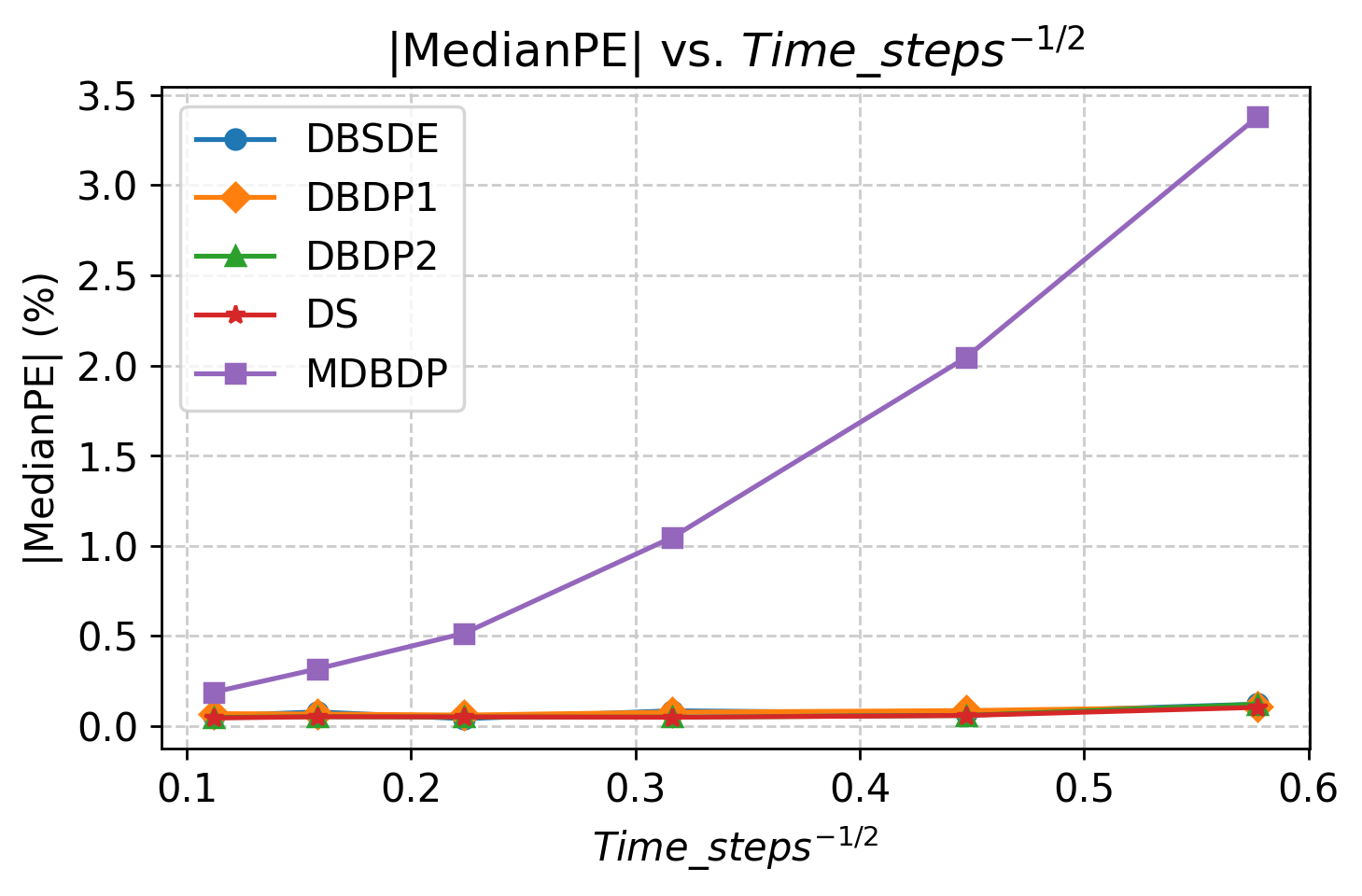}
  \vspace*{-4mm}
  \caption{Experiment 2 (Time steps). MDBDP produces substantially higher errors as the number of time steps decreases (rightward in the graph). Note that all algorithms but MDBDP almost overlap. }
  \label{fig:NvM}
\end{figure}

\begin{figure}[t]
 \centering
  \includegraphics[width=8.5cm]{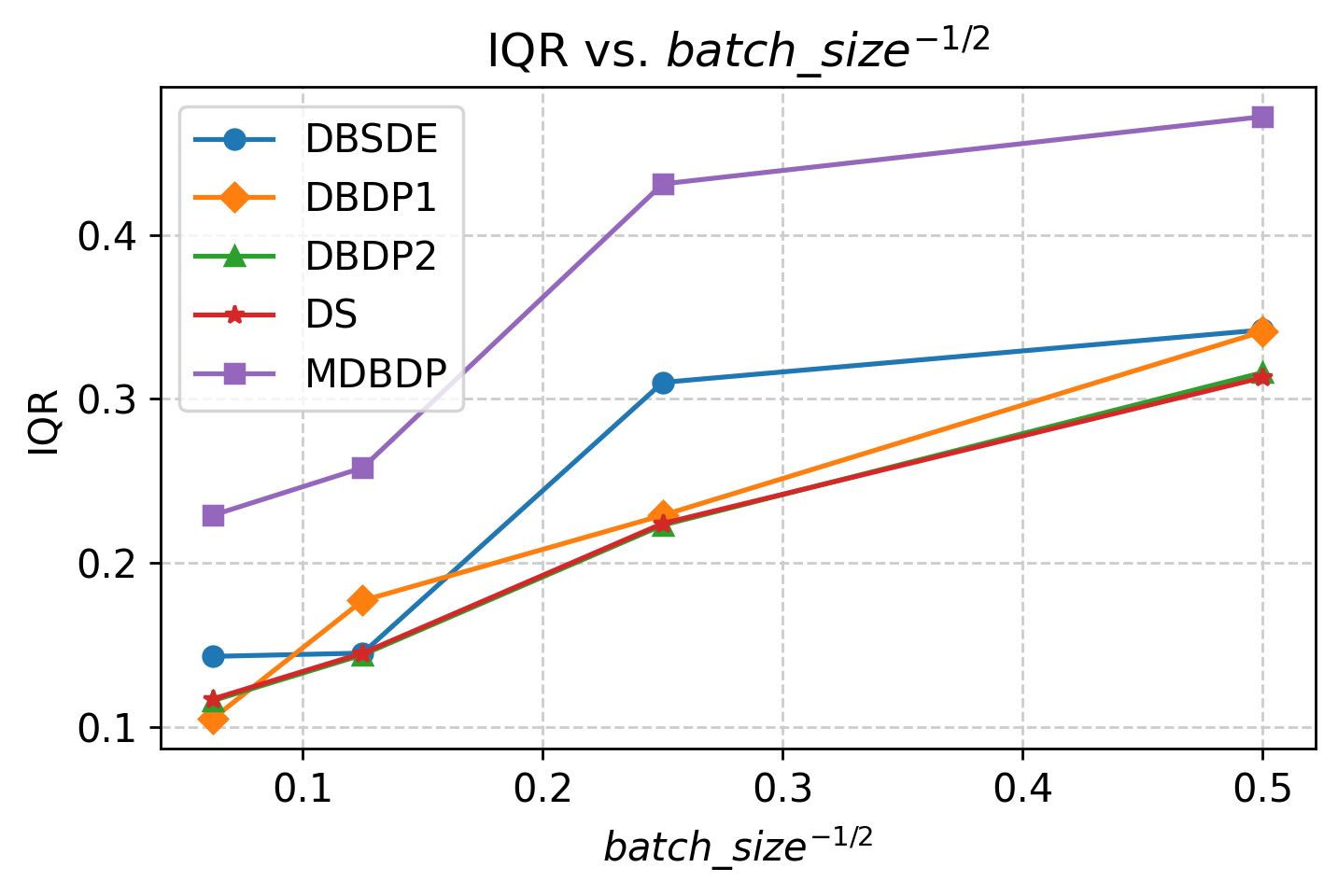}
  \vspace*{-4mm}
  \caption{Experiment 2 (Batch size). Special attention goes to this figure since the measurement of error is the interquartile range not MedianPE. }
  \label{fig:bvI}
\end{figure}

\begin{figure}[t]
 \centering
  \includegraphics[width=8.5cm]{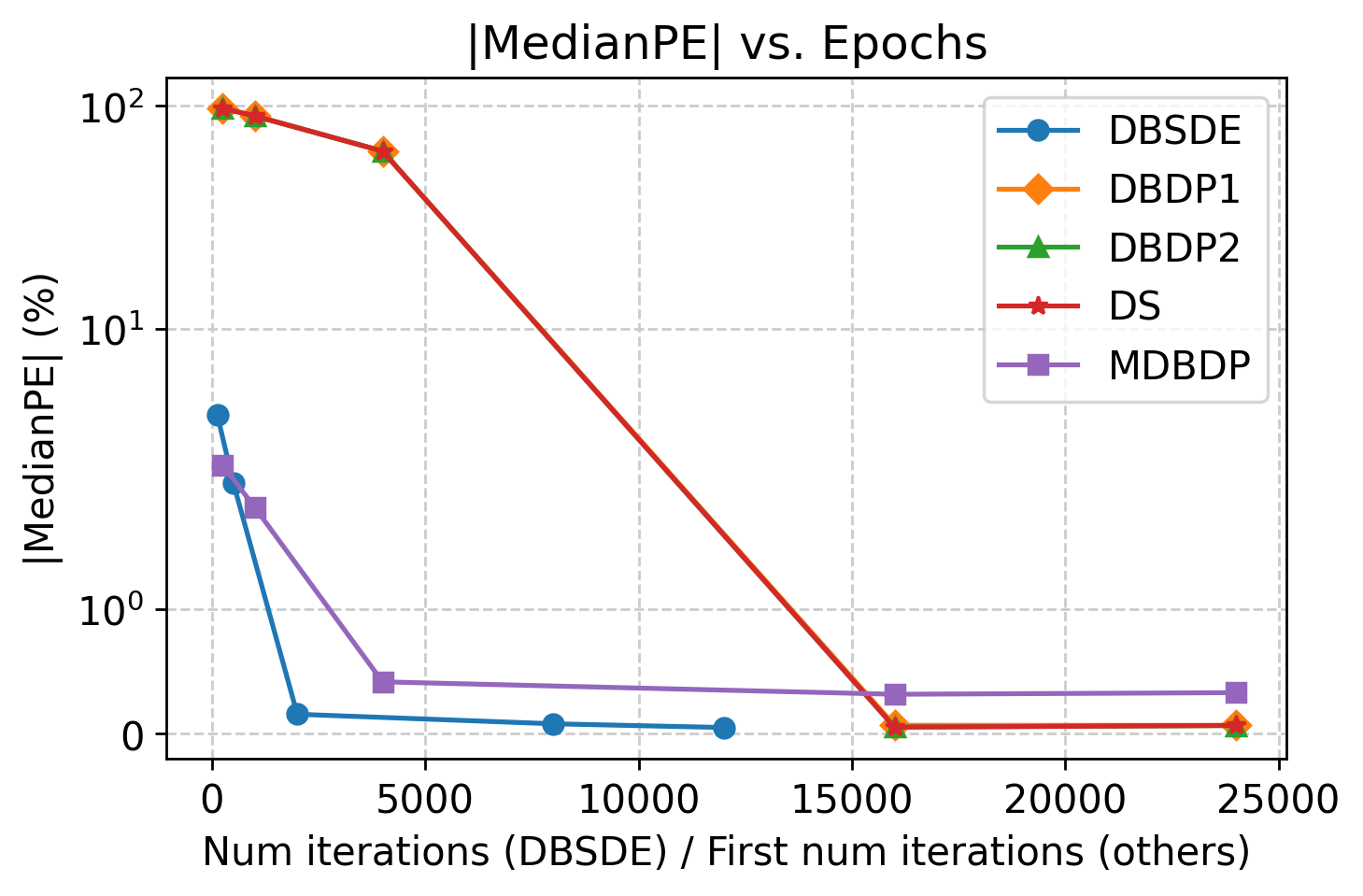}
  \vspace*{-4mm}
  \caption{Experiment 3 (Epochs). This figure incorporates the results from all algorithms for illustration. However, the results from the forward scheme (DBSDE) cannot be compared with those from the backward schemes (others). In case of insufficient epochs, MDBDP's errors are lower than all of the other backward scheme algorithms. }
  \label{fig:numvM}
\end{figure}

Every single setting described earlier is executed 100 times with all algorithms, and the results are reported in Quartile along with the Monte Carlo solutions. The Monte Carlo solutions are deemed as the accurate representation of true price values. Each Monte Carlo solution is obtained by $10^6$ asset price simulations where the time interval is divided into $10^4$ steps. The numerical results of all experiments are 
plotted in \Cref{fig:TvM} to \Cref{fig:numvM}, respectively. Every setting in Experiment 1 is paired with a Monte Carlo solution. On the other hand, in Experiments 2 and 3, the Monte Carlo solution is not given because they all represent the standard setting, which has the Monte Carlo option price of $94.933$. Besides, either \textit{MedianPE} or \textit{IQR} is reported.  MedianPE, or Median Percentage Error, refers to the percentage error of the median in comparison to the Monte Carlo value, whereas IQR refers to the interquartile range – which is calculated by subtracting Q1 from Q3. All factors but batch size have MedianPE as the measurement of error, and its absolute values are plotted in the figures. In the case of batch size, IQR is chosen instead for the fact that the spread of solution distribution rather than its median is influenced by the choice of batch size. 

\section{Discussion}\label{sec:discussion}
 
This section makes empirical analysis and discussion on the obtained results reported in \cref{Numexp:result}. Accordingly, the hypotheses about sources of errors illustrated in \cref{sec:source} will be assessed. This section is also divided into four subsections. The first three subsections represent each of the error types: options' and underlying assets' uncertainties, asset models, and optimization. The last subsection provides a unified comparison among the tested algorithms and our practical guidance on using deep PDE solvers. 

\subsection{Errors inherent in options' and underlying assets' uncertainties}

Regarding time to expiration as in \Cref{fig:TvM}, DBDP1, DBDP2, and DS depict not only an increment in error with lengthier times but, notably, glaring errors appear at some points beyond $T=1.0$. On the other hand, MDBDP and DBSDE demonstrate greater resilience and consistency even as they cause more error when the time to expiration increases, with MDBDP's error growing faster than DBSDE's. 



In terms of moneyness, the experimental results illustrate that options deeper in the money are associated with less pricing error. DBSDE, DBDP1, DBDP2, and DS all showed a decline in MedianPE as moneyness increases. Throughout the region, DBDP2 produced the lowest error but not significantly lower than DS. By contrast, errors produced by MDBDP were the highest in any tested moneyness, but MDBDP demonstrated a more consistent performance over the studied range. Its small fluctuation, from $0.276 \%$ to $0.333 \%$ as shown in \Cref{fig:MvM}, is attributed to inexplicable factors. 

Generally, as the long-term variance increases, MedianPE shifts downward. Considering the fact that all experiments were run with the initial variance of 0.1, we found that the obtained solutions tend to shift toward the correct prices of options with the final variance of 0.1 because the initial variance was not updated to the defined long-term variance soon enough. In other words, the price of an option where the initial variance was lower than 0.1 was predicted to be higher than its accurate value because its expected variance might not be updated in a timely manner to center around the lower long-term variance. In DBDP2 and DS, the contrary effect happened to be the case: the prices of options where the initial variance was higher than 0.1 were predicted to be lower than their accurate value as their expected variance might not get updated to the higher long-term variance. Nevertheless, the above explanation could not explain the results from DBSDE and MDBDP, and we must attribute the cause of errors in those two algorithms to unexplainable factors. 

\subsection{Errors by asset models}




We examined how changing the number of time intervals and batch sizes affected the final errors. The relationship between $1 / \sqrt{\text{batch size}}$ and the results' spread, as determined by the IQR, was clearly proportional for the DBDP1, DBDP2, and DS algorithms. It is firmly illustrated that larger batch sizes produce more consistent results centered around its median. This is consistent with the idea that a small batch size might not fairly reflect the price distribution, which could result in inconsistent estimated option prices. Since there are confirmed advantages to enlarging the batch, it is strongly recommended that batch size in any algorithm is as large as possible. This is especially advised in environments equipped with GPUs where it is unlikely that a larger batch size will hold back computation.

There exists a particularly strong linear relationship between $1 / \sqrt{\text{time steps}}$ and error. This confirms our hypothesis that time discretization introduces errors in asset price sampling by suggesting that the error in option pricing tends to decrease as the number of time steps increases. This pattern is most noticeable in MDBDP, where MedianPE escalated by fewer time steps. On the other hand, the other algorithms appear to be affected by other factors, leading to inconsistent and fluctuating outcomes. Besides, although the execution time may not be linearly affected by an increase in the number of time steps, the process may not be as simple as augmenting the batch size. While raising the number of time steps to the greatest extent possible is most preferable, it is recommended to implement at least 20 to 40 time steps, irrespective of the algorithm or time to expiration. 




\subsection{Optimization errors}




The findings on how the number of iterations affected the accuracy highlight the importance of choosing the right number of iterations. MedianPE strictly decreases as the number of iterations increases. Marked by lowest errors, the best performer was DBSDE, which incorporates the acceleration scheme \cite{naito2020acceleration} in the beginning. One persistent challenge is the absence of a conclusive approach to ascertain the optimal number of iterations, which still requires a trade-off between computational efficiency and solution accuracy. Moreover, the optimal number is subject to a radical change by various factors, e.g., learning rate.

\subsection{Unified comparison}

When comparing the performance of different deep PDE solvers in different option pricing scenarios, a significant difference in effectiveness is revealed by changing the specified parameters, particularly the time to expiration and long-term variance. The general ability of DBDP2 and DS to achieve lower MedianPE breaks down when the time to expiration is either too short ($\leq 0.25$) or too long ($\geq 1.25$). In these particular cases, DBSDE is the most robust and accurate model, as shown by the MeanPE results in \Cref{fig:TvM}. Validated by \Cref{fig:vvM}, this resilience of DBSDE further extrapolates to scenarios with high mean variance ($\geq 0.14$), which seriously compromises the accuracy of DBDP2 and DS, offering DBSDE as a comparative advantage in these conditions. 

When this analysis is applied practically, traders must carefully select their algorithms based on particular market conditions and computational limitations. In particular, DBSDE is a strong solution that reduces the errors shown in DBDP2 and DS for options with extended or truncated expiration timelines or those with high mean variance. 
Additionally, even though MDBDP appears competent, its computational cost is not worthwhile in the financial context where speed is a priority. Therefore, we recommend that the pricing of European-style options with many underlying assets (solving high-dimensional PDEs) should be done with DBSDE. On the other hand, for other option pricing problems where DBSDE cannot be applied, which include but are not limited to pricing American-style options\footnote{
    To explore how (M)DBDP is extended to solve variational inequalities, including the pricing of American-style options, the reader is referred to \cite{hure2020backward}.
}, we suggest that they are solved by MDBDP, which possesses an ability to return the solution path $u(t_i, \cdot)$ at any time step $i$ like other backward schemes. 

Regarding our recommendation on settings, it is advisable to always use configurations that at least achieve the best performance within the scope of this paper, especially the batch size, for financial applications where the strictest error reduction measures are crucial. In a real situation, an accurate solution is expected from a single run, and it is necessary to narrow down the spread of an obtained solution. Therefore, we recommend applying the batch size with a minimum of 256 for better consistency and lower variability. 

Though these suggestions have a solid empirical basis, it is important to recognize that their applicability is limited to the parameter space that has been studied, which is around the standard parameters in the Heston model. Options with parameters that deviate markedly from our experimental setup might show patterns and behaviors that our current model and comprehension are unable to capture or anticipate. Option PDEs are subject to additional value adjustment terms (known as xVAs), which would alter our findings to some degree if considered.


\section{Conclusion}\label{sec:conclusion}
This paper systematically investigated the error sources in the deep PDE/BSDE solvers on their use in option pricing. Five deep PDE solvers, including DBSDE, DBDP1, DBDP2, DS, and MDBDP, were utilized to solve the Heston PDE and compute the option price in a specified setting. Through a series of devised numerical experiments, insights into errors inherent in options' and underlying assets' uncertainties, asset models, and optimization were empirically explored. The results revealed different advantages of the tested algorithms, and they could confirm relationships between the error and some factors varied in the experiments.

Following the results, we subsequently provided practical advice on how to utilize deep PDE solvers in an efficient manner. In addition, we could solidly reveal the harmful effects of inadequate numbers of time steps, iterations, etc. In addition, we recommended how solvers and settings should be chosen for resilience against inaccuracies. We also discussed the inherent limitations of our experimental environments, which, while informative, do not fully capture the complex nature of options pricing. 


\bibliographystyle{IEEEtran}
\bibliography{main}

\clearpage
\onecolumn
\section*{Appendix: Complete tables of the experimental results}

In this supplementary material, we provide the complete tables of the experimental results discussed in \Cref{sec:experiment}.
\Cref{tab:11} to \Cref{tab:31} correspond to \Cref{fig:TvM} to \Cref{fig:numvM}, respectively.

\begin{table*}[h!]
  \centering
  \caption{Experiment 1 (Time to expiration)}
\begin{tabular}{|c|c|c|c|c|c|c|c|}
\hline  
Time to expiration
& $3 / 12$ & $6 / 12$ & $9 / 12$ & $12 / 12$ & $15 / 12$ & $18 / 12$ & $21 / 12$ \\
\hline Monte Carlo  & 50.857 & 67.882 & 82.122 & 94.933 & 106.836 & 118.101 & 128.883 \\
\hline DBSDE $25 \%$ & 50.830 & 67.856 & 82.100 & 94.919 & 106.823 & 118.096 & 128.887 \\
DBSDE Median & 50.884 & 67.925 & 82.187 & 95.007 & 106.925 & 118.203 & 129.008 \\
DBSDE $75 \%$ & 50.942 & 67.989 & 82.241 & 95.064 & 106.999 & 118.278 & 129.080 \\
DBSDE MedianPE & \textbf{0.053\%} & $0.064 \%$ & $0.079 \%$ & $0.078 \%$ & \textbf{0.083\%} & \textbf{0.086\%} & \textbf{0.097\%} \\
\hline DBDP1 25\% & 50.820 & 67.845 & 82.079 & 94.905 & 106.612 & 116.235 & 122.475 \\
DBDP1 Median & 50.886 & 67.926 & 82.185 & 94.996 & 106.698 & 116.311 & 122.535 \\
DBDP1 75\% & 50.937 & 68.000 & 82.259 & 95.082 & 106.776 & 116.374 & 122.580 \\
DBDP1 MedianPE & $0.056 \%$ & $0.065 \%$ & $0.076 \%$ & $0.066 \%$ & $-0.130 \%$ & $-1.515 \%$ & $-4.926 \%$ \\
\hline DBDP2 25\% & 50.845 & 67.869 & 82.110 & 94.929 & 106.626 & 116.253 & 122.494 \\
DBDP2 Median & 50.891 & 67.921 & 82.171 & 94.981 & 106.677 & 116.299 & 122.518 \\
DBDP2 75\% & 50.943 & 68.002 & 82.251 & 95.073 & 106.784 & 116.383 & 122.587 \\
DBDP2 MedianPE & $0.066 \%$ & $0.057 \%$ & $0.059 \%$ & \textbf{0.051\%} & $-0.149 \%$ & $-1.526 \%$ & $-4.938 \%$ \\
\hline DS $ 25 \%$ & 50.848 & 67.871 & 82.111 & 94.929 & 106.627 & 116.255 & 122.492 \\
DS Median & 50.886 & 67.919 & 82.170 & 94.982 & 106.677 & 116.299 & 122.520 \\
DS $75 \%$ & 50.944 & 68.001 & 82.252 & 95.074 & 106.784 & 116.383 & 122.591 \\
DS MedianPE & $0.057 \%$ & \textbf{0.054\%} & \textbf{0.058\%} & \textbf{0.051\%} & $-0.149 \%$ & $-1.525 \%$ & $-4.937 \%$ \\
\hline MDBDP $25 \%$ & 50.828 & 67.930 & 82.247 & 95.089 & 107.087 & 118.515 & 129.335 \\
MDBDP Median & 50.909 & 68.010 & 82.340 & 95.233 & 107.245 & 118.626 & 129.529 \\
MDBDP $75 \%$ & 50.970 & 68.095 & 82.439 & 95.347 & 107.376 & 118.773 & 129.700 \\
MDBDP MedianPE & $0.102 \%$ & $0.189 \%$ & $0.265 \%$ & $0.316 \%$ & $0.382 \%$ & $0.445 \%$ & $0.501 \%$ \\
\hline
\end{tabular}
\label{tab:11}
\end{table*}

\begin{table*}[h!]
  \centering
  \caption{Experiment 1 (Moneyness)}
\begin{tabular}{|c|c|c|c|c|c|}
\hline Moneyness & 0.9 & 1 & 1.1 & 1.2 & 1.3 \\
\hline Monte Carlo & 68.510 & 79.079 & 87.727 & 94.933 & 101.030 \\
\hline DBSDE 25\% & 68.495 & 79.065 & 87.713 & 94.919 & 101.017 \\
DBSDE Median & 68.583 & 79.153 & 87.800 & 95.007 & 101.105 \\
DBSDE $75 \%$ & 68.640 & 79.209 & 87.858 & 95.064 & 101.163 \\
DBSDE MedianPE & $0.106 \%$ & $0.093 \%$ & $0.084 \%$ & $0.078 \%$ & $0.073 \%$ \\
\hline DBDP1 25\% & 68.476 & 79.049 & 87.695 & 94.905 & 100.978 \\
DBDP1 Median & 68.586 & 79.153 & 87.790 & 94.996 & 101.068 \\
DBDP1 75\% & 68.675 & 79.236 & 87.890 & 95.082 & 101.149 \\
DBDP1 MedianPE & $0.111 \%$ & $0.094 \%$ & $0.073 \%$ & $0.066 \%$ & $0.037 \%$ \\
\hline DBDP2 25\% & 68.508 & 79.078 & 87.725 & 94.929 & 101.006 \\
DBDP2 Median & 68.565 & 79.135 & 87.781 & 94.981 & 101.052 \\
DBDP2 75\% & 68.664 & 79.233 & 87.878 & 95.073 & 101.144 \\
DBDP2 MedianPE & \textbf{0.081\%} & \textbf{0.070\%} & \textbf{0.062\%} & \textbf{0.051\%} & \textbf{0.021\%} \\
\hline DS $25 \%$ & 68.507 & 79.076 & 87.723 & 94.929 & 101.005 \\
DS Median & 68.566 & 79.134 & 87.781 & 94.982 & 101.053 \\
DS $75 \%$ & 68.665 & 79.234 & 87.877 & 95.074 & 101.144 \\
DS MedianPE & $0.082 \%$ & \textbf{0.070\%} & $0.063 \%$ & \textbf{0.051\%} & $0.023 \%$ \\
\hline MDBDP $25 \%$ & 68.616 & 79.199 & 87.907 & 95.089 & 101.254 \\
MDBDP Median & 68.711 & 79.297 & 88.019 & 95.233 & 101.338 \\
MDBDP $75 \%$ & 68.865 & 79.466 & 88.104 & 95.347 & 101.465 \\
MDBDP MedianPE & $0.293 \%$ & $0.276 \%$ & $0.333 \%$ & $0.316 \%$ & $0.305 \%$ \\
\hline
\end{tabular}
\label{tab:12}
\end{table*}

\begin{table*}[h!]
  \centering
  \caption{Experiment 1 (Long-term variance)}
\begin{tabular}{|c|c|c|c|c|c|}
\hline Long-term variance & 0.06 & 0.08 & 0.10 & 0.12 & 0.14 \\
\hline Monte Carlo & 84.287 & 89.717 & 94.933 & 99.971 & 104.857 \\
\hline DBSDE 25\% & 84.326 & 89.730 & 94.919 & 99.928 & 104.785 \\
DBSDE Median & 84.414 & 89.819 & 95.007 & 100.018 & 104.879 \\
DBSDE 75\% & 84.488 & 89.883 & 95.064 & 100.082 & 104.947 \\
DBSDE MedianPE & $0.151 \%$ & $0.114 \%$ & $0.078 \%$ & $0.048 \%$ & \textbf{0.021\%} \\
\hline DBDP1 25\% & 84.327 & 89.723 & 94.905 & 99.891 & 104.662 \\
DBDP1 Median & 84.440 & 89.818 & 94.996 & 99.984 & 104.767 \\
DBDP1 75\% & 84.505 & 89.900 & 95.082 & 100.068 & 104.858 \\
DBDP1 MedianPE & $0.181 \%$ & $0.112 \%$ & $0.066 \%$ & $0.013 \%$ & $-0.086 \%$ \\
\hline DBDP2 25\% & 84.360 & 89.750 & 94.929 & 99.920 & 104.701 \\
DBDP2 Median & 84.400 & 89.800 & 94.981 & 99.966 & 104.747 \\
DBDP2 75\% & 84.490 & 89.894 & 95.073 & 100.064 & 104.853 \\
DBDP2 MedianPE & \textbf{0.134\%} & \textbf{0.092\%} & \textbf{0.051\%} & $-0.005 \%$ & $-0.105 \%$ \\
\hline DS 25\% & 84.360 & 89.750 & 94.929 & 99.920 & 104.701 \\
DS Median & 84.399 & 89.800 & 94.982 & 99.969 & 104.749 \\
DS 75\% & 84.490 & 89.895 & 95.074 & 100.063 & 104.851 \\
DS MedianPE & \textbf{0.134\%} & \textbf{0.092\%} & \textbf{0.051\%} & \textbf{--0.002\%} & $-0.103 \%$ \\
\hline MDBDP 25\% & 84.508 & 89.917 & 95.089 & 100.133 & 105.037 \\
MDBDP Median & 84.606 & 90.006 & 95.233 & 100.244 & 105.144 \\
MDBDP 75\% & 84.714 & 90.134 & 95.347 & 100.356 & 105.237 \\
MDBDP MedianPE & $0.379 \%$ & $0.322 \%$ & $0.316 \%$ & $0.273 \%$ & $0.274 \%$ \\
\hline
\end{tabular}
\label{tab:13}
\end{table*}

\begin{table*}[h!]
  \centering
  \caption{Experiment 2 (Time steps)}

\begin{tabular}{|c|c|c|c|c|c|c|}
\hline Time steps & 3 & 5 & 10 & 20 & 40 & 80 \\
\hline DBSDE $25 \%$ & 94.905 & 94.908 & 94.911 & 94.881 & 94.919 & 94.910 \\
DBSDE Median & 95.048 & 94.999 & 95.013 & 94.971 & 95.007 & 94.985 \\
DBSDE 75\% & 95.153 & 95.144 & 95.102 & 95.058 & 95.064 & 95.070 \\
DBSDE MedianPE & $0.121 \%$ & $0.070 \%$ & $0.084 \%$ & \textbf{0.040\%} & $0.078 \%$ & $0.055 \%$ \\
\hline DBDP1 25\% & 94.942 & 94.907 & 94.932 & 94.906 & 94.905 & 94.901 \\
DBDP1 Median & 95.034 & 95.014 & 95.005 & 94.989 & 94.996 & 94.997 \\
DBDP1 75\% & 95.110 & 95.109 & 95.093 & 95.071 & 95.082 & 95.059 \\
DBDP1 MedianPE & $0.107 \%$ & $0.085 \%$ & $0.076 \%$ & $0.060 \%$ & $0.066 \%$ & $0.068 \%$ \\
\hline DBDP2 25\% & 94.979 & 94.887 & 94.907 & 94.900 & 94.929 & 94.891 \\
DBDP2 Median & 95.046 & 94.988 & 94.979 & 94.980 & 94.981 & 94.978 \\
DBDP2 75\% & 95.120 & 95.075 & 95.081 & 95.051 & 95.073 & 95.064 \\
DBDP2 MedianPE & $0.120 \%$ & \textbf{0.058\%} & $0.049 \%$ & $0.049 \%$ & \textbf{0.051\%} & $0.047 \%$ \\
\hline DS $25 \%$ & 94.958 & 94.887 & 94.907 & 94.900 & 94.929 & 94.889 \\
DS Median & 95.030 & 94.988 & 94.978 & 94.980 & 94.982 & 94.974 \\
DS $75 \%$ & 95.094 & 95.075 & 95.080 & 95.051 & 95.074 & 95.070 \\
DS MedianPE & \textbf{0.103\%} & \textbf{0.058\%} & \textbf{0.048\%} & $0.050 \%$ & \textbf{0.051\%} & \textbf{0.044\%} \\
\hline MDBDP 25\% & 98.008 & 96.769 & 95.828 & 95.307 & 95.089 & 94.993 \\
MDBDP Median & 98.141 & 96.872 & 95.925 & 95.421 & 95.233 & 95.110 \\
MDBDP $75 \%$ & 98.320 & 96.988 & 96.020 & 95.556 & 95.347 & 95.228 \\
MDBDP MedianPE & $3.380 \%$ & $2.043 \%$ & $1.045 \%$ & $0.514 \%$ & $0.316 \%$ & $0.187 \%$ \\
\hline
\end{tabular}

\label{tab:21}
\end{table*}

\begin{table*}[h!]
  \centering
  \caption{Experiment 2 (Batch size)}
\begin{tabular}{|c|c|c|c|c|}
\hline Batch size & 4 & 16 & 64 & 256 \\
\hline DBSDE 25\% & 94.852 & 94.852 & 94.919 & 94.932 \\
DBSDE Median & 95.078 & 95.034 & 95.007 & 94.992 \\
DBSDE 75\% & 95.194 & 95.162 & 95.064 & 95.075 \\
DBSDE IQR & 0.342 & 0.310 & 0.145 & 0.143 \\
\hline DBDP1 25\% & 94.036 & 94.796 & 94.905 & 94.935 \\
DBDP1 Median & 94.266 & 94.914 & 94.996 & 94.985 \\
DBDP1 75\% & 94.378 & 95.025 & 95.082 & 95.039 \\
DBDP1 IQR & 0.341 & 0.229 & 0.177 & \textbf{0.105} \\
\hline DBDP2 25\% & 94.038 & 94.820 & 94.929 & 94.942 \\
DBDP2 Median & 94.176 & 94.913 & 94.981 & 94.997 \\
DBDP2 75\% & 94.354 & 95.042 & 95.073 & 95.058 \\
DBDP2 IQR & 0.316 & \textbf{0.223} & \textbf{0.144} & 0.116 \\
\hline DS 25\% & 94.039 & 94.819 & 94.929 & 94.941 \\
DS Median & 94.177 & 94.914 & 94.982 & 94.997 \\
DS $75 \%$ & 94.352 & 95.042 & 95.074 & 95.058 \\
DS IQR & \textbf{0.313} & 0.224 & 0.145 & 0.117 \\
\hline MDBDP 25\% & 94.990 & 94.985 & 95.089 & 95.084 \\
MDBDP Median & 95.284 & 95.211 & 95.233 & 95.187 \\
MDBDP 75\% & 95.462 & 95.416 & 95.347 & 95.312 \\
MDBDP IQR & 0.472 & 0.431 & 0.258 & 0.229 \\
\hline
\end{tabular}
\label{tab:22}
\end{table*}

\begin{table*}[h!]
  \centering
  \caption{Experiment 3 (Epochs)}
\begin{tabular}{|c|c|c|c|c|c|}
\hline Epochs & 125 & 500 & 2000 & 8000 & 12000 \\
\hline DBSDE 25\% & 98.320 & 96.543 & 94.973 & 94.919 & 94.901 \\
DBSDE Median & 98.866 & 96.883 & 95.080 & 95.007 & 94.978 \\
DBSDE $75 \%$ & 99.600 & 97.419 & 95.168 & 95.064 & 95.097 \\
 DBSDE MedianPE & \textbf{4.143\%} & \textbf{2.054\%} & \textbf{0.155\%} & \textbf{0.078\%} & \textbf{0.048\%} \\
\hline First-time-step epochs & 250 & 1000 & 4000 & 16000 & 24000 \\
\hline DBDP1 25\% & 2.368 & 9.307 & 35.389 & 94.905 & 94.897 \\
DBDP1 Median & 2.377 & 9.318 & 35.403 & 94.996 & 94.994 \\
DBDP1 75\% & 2.383 & 9.330 & 35.414 & 95.082 & 95.074 \\
DBDP1 MedianPE & $-97.497 \%$ & $-90.185 \%$ & $-62.708 \%$ & $0.066 \%$ & $0.065 \%$ \\
\hline DBDP2 25\% & 2.373 & 9.309 & 35.396 & 94.929 & 94.900 \\
DBDP2 Median & 2.376 & 9.316 & 35.406 & 94.981 & 94.993 \\
DBDP2 $75 \%$ & 2.379 & 9.324 & 35.418 & 95.073 & 95.116 \\
DBDP2 MedianPE & $-97.497 \%$ & $-90.187 \%$ & $-62.705 \%$ & \textbf{0.051\%} & \textbf{0.064 \%} \\
\hline DS 25\% & $\begin{array}{l}2.371 \\
\end{array}$ & 9.309 & 35.396 & 94.929 & 94.900 \\
DS Median & 2.376 & 9.316 & 35.406 & 94.982 & 94.994 \\
DS $75 \%$ & 2.380 & 9.323 & 35.417 & 95.074 & 95.116 \\
DS MedianPE & $-97.497 \%$ & $-90.186 \%$ & $-62.704 \%$ & \textbf{0.051\%} & $0.065 \%$ \\
\hline MDBDP 25\% & 97.176 & 95.485 & 95.217 & 95.089 & 95.132 \\
MDBDP Median & 97.274 & 96.664 & 95.328 & 95.233 & 95.245 \\
MDBDP 75\% & 97.391 & 97.018 & 95.448 & 95.347 & 95.367 \\
MDBDP MedianPE & \textbf{2.466\%} & \textbf{1.824\%} & \textbf{0.416 \%} & $0.316 \%$ & $0.329 \%$ \\
\hline
\end{tabular}
\\
Because forward and backward schemes cannot be compared against each other, the top performers are separately embolden. 
\label{tab:31}
\end{table*}

\end{document}